# Enhancing power of rare variant association test by Zoom-Focus Algorithm (ZFA) to locate optimal testing region

Maggie Haitian Wang[1+*], Haoyi Weng[1+], Rui Sun[1+], Benny Chung-Ying Zee[1*]

[1]Division of Biostatistics, JC School of Public Health and Primary Care, the Chinese University of Hong Kong, Shatin, N.T., Hong Kong SAR; CUHK Shenzhen Research Institute, Shenzhen, China.

[+] These authors contribute equally to this work
[*]To whom correspondence should be addressed.

**ABSTRACT**

**Motivation:** Exome or targeted sequencing data exerts analytical challenge to test single nucleotide polymorphisms (SNPs) with extremely small minor allele frequency (MAF). Various rare variant tests were proposed to increase power by aggregating SNPs within a fixed genomic region, such as a gene or pathway. However, a gene could contain from several to thousands of markers, and not all of them may be related to the phenotype. Combining functional and non-functional SNPs in arbitrary genomic region could impair the testing power.

**Results:** We propose a Zoom-Focus algorithm (ZFA) to locate the optimal testing region within a given genomic region, as a wrapper function to be applied in conjunction with rare variant association tests. In the first Zooming step, a given genomic region is partitioned by order of two, and the best partition is located within all partition levels. In the next Focusing step, boundaries of the zoomed region are refined. Simulation studies showed that ZFA substantially enhanced the statistical power of rare variant tests by over 10 folds, including the WSS, SKAT and W-test. The algorithm is applied on real exome sequencing data of hypertensive disorder, and identified biologically relevant genetic markers to metabolic disorder that are undiscoverable by testing using full gene. The proposed algorithm is an efficient and powerful tool to increase the effectiveness of rare variant association tests for exome sequencing datasets of complex disorder.

**Contact:** maggiew@cuhk.edu.hk

# 1 INTRODUCTION

The increasing depth and coverage of DNA sequencing technology have made possible the accumulation of human whole genome sequencing data. National-level efforts are undergone to collect big cohort deep sequencing data towards better knowledge of Mendelian and complex disorders (Ashley, 2015; Auffray, *et al*., 2016; Cyranoski, 2016; Jameson and Longo, 2015). These efforts press the demand of powerful and efficient methods for medical and clinical inference from the data. The challenge in analyzing the exome sequencing data set, besides the multiple-testing issue arising from the high dimensionality, centers on the extreme low allele frequencies - Classical statistical tests lose power on single nucleotide polymorphisms (SNPs) with small variances. In whole exome sequencing data, over 99% of the SNPs have minor allele frequency (MAF) below 1% (Consortium, 2015). A number of rare variant association tests have been proposed, which either pull adjacent SNPs together to increase minor allele frequencies and up-weight the minor allele (burden tests) (Li and Leal, 2008; Liu and Leal, 2010; Madsen and Browning, 2009), or apply a linear mixed model on a certain genomic region to improve power (variance component tests) *(Lee, et al., 2012; Neale, et al., 2011; Wu, et al*., 2011). For most rare variant methods, a fixed genomic region for testing is assumed, such as a gene or a fixed window. However, in real data application, a gene may contain from several to thousands of SNPs. Directly applying rare variant association tests based on a gene or a fixed window may introduce a large number of un-necessary noises that could impair testing power (Santorico and Hendricks, 2016). Furthermore, many of the exome sequencing data have unknown gene functions and are difficult to pool without prior information. It is desirable to optimize the collapsing region such that minimum noise is included and power of the test can be enhanced.

In this paper, we propose a Zoom-Focus algorithm (ZFA) to optimize testing region for rare variant association tests. The algorithm can be applied as a wrapper function to existing tests and can increase the test power by over 10-fold in simulations studies. The algorithm takes two main steps, Zooming and Focusing. In the Zooming step, a fix genomic region is partitioned by an order of two, and search is conducted through all partition levels to identify the region with maximum information, evaluated by the smallest association p-value of that partition. Based on the best partition identified by Zooming, a next Focusing step is used to optimize the region's boundary, by adding or reducing adjacent SNPs around the boundaries. Simulation studies for various genetic scenarios demonstrated that the Zoom-Focus algorithm could substantially enhance the statistical power of different rare variant methods, including the WSS, SKAT and W-test. The ZFA is applied on real exome sequencing data of hypertensive disorder and identified biologically relevant genetic markers that are undiscoverable by un-optimized testing regions.

## 2 METHODS

### 2.1 The Zoom-Focus Algorithm (ZFA)

The Zoom-Focus Algorithm is best described through an example. Suppose a gene or fixed windows contains 64 SNPs, among which 8 SNPs are causal. For the simplest scenario, the 8 causal markers cluster together, as shown in the red region in Diagram 1. (Other causal marker distributions are provided in the simulation study.) The ZFA first performs an exhaustive search in all possible binary partitions of the initial region to locate the best partition (Zooming), and then adjust the boundary of the zoomed region by considering increment or decrement of the bounds (Focusing) (Diagram 1).

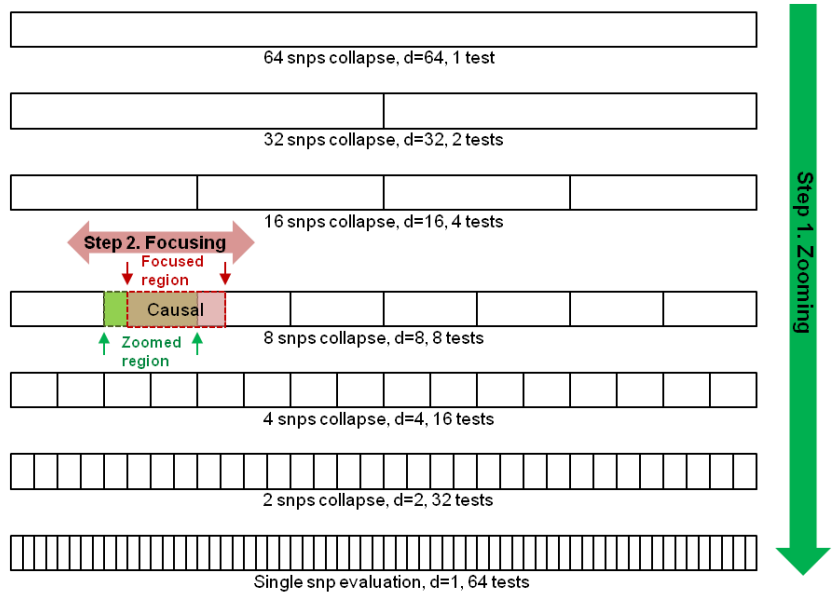

**Diagram 1.** The Zoom-Focus Algorithm (ZFA)
Legend: The ZFA first performs an exhaustive search in all possible binary partitions in a given genomic region (Step1: Zooming); then adjust the boundary of the zoomed region by considering increment or decrement of the bounds (Step2: Focusing).

We will first define some notations, and then introduce the algorithm. Let $P$ be the total number of SNPs in a fixed window; $R$ is the maximum order of binary partition for $P$, and $R = \arg\max\{r: 2^{r-1} \leq P, r = 1, 2, \ldots\}$; $r$ is the order of partition. At a certain $r$, the number of partitioned regions is $n_r = 2^{r-1}$, and the size of partition $d$ is the number of SNPs in a partitioned region, $d = P/n_r = P/2^{r-1}$. A higher order gives a smaller partitioned region size. $c$ is the index for the $c^{th}$ partition, $c = 1, \ldots, n_r$. For the causal region in Diagram 1: $P = 64$, the causal region is located at $4^{th}$ order of binary division and the $2^{nd}$ partitioned region ($c=2$), thus $r = 4$, and $d = 8$. We denote $\phi(d, c) = \{x_{cj}, j=1, \ldots, d\}$ as the region that contains SNPs in a partition size $d$ and index $c$, and the causal region can be denoted as $\phi(8, 2)$. Let $f(\cdot)$ be the Bonferroni corrected $p$-value calculated by rare variant method $F(\cdot)$ measured on region $\phi(d, c)$:

$$f(d, c) = n_r \cdot F[\phi(d, c)], \qquad \text{Eq 1.}$$

where $n_r$ is the number of partitions for Bonferroni correction of multiple tests. It ensures p-values at the different partition order can be compared. Therefore, we have:

**Step 1: Zooming.** Search all binary partitions of a given initial region, for an optimal partition of size $d$ and index $c$: $\phi(d, c)$, such that $f(d, c)$ is minimized:

$$(\hat{d}, \hat{c}) = \arg\min\{ f(d, c), r = 0,\ldots,R; c = 1 \text{ to } n_r \} \qquad \text{Eq 2.}$$

**Step 2: Focusing.** Refine the boundary of $(\hat{d}, \hat{c})$ by extending both lower and upper bounds, outwardly by d/2, and inwardly by d/4. Let LB denotes the lower bound of $(\hat{d}, \hat{c})$, and UB denotes the upper bound of $(\hat{d}, \hat{c})$. The refined lower bound ($LB_f$) and upper bound ($UB_f$) are:

$$LB_f = \arg\min\{ f(LB', UB), LB' = LB + i\,; i = [-d/2, d/4] \}, \text{ and} \qquad \text{Eq 3.}$$
$$UB_f = \arg\min\{ f(LB_f, UB'), UB' = UB + i\,; i = [-d/4, d/2] \}.$$

In sum, the Zooming step locates the best partition from global search, and the Focusing step refines the boundary of the zoomed region from local linear optimization. The terms are motivated from optical zooming lens.

## 2.2 Computation complexity

In the Zooming step, the maximum number of tests $T(P)$ in a gene with $P$ number of SNPs is:

$$T(P) = \sum_{r=0}^{R} 2^r = 2^0 + 2^1 + \ldots + 2^R = \frac{1 - 2^{R+1}}{1 - 2} = 2^{R+1} - 1 \approx 2^{1 + \log_2(P) + 1} - 1 = 2P - 1.$$

In the Focusing step, the worst-case scenario contains $P/2$ calculations. Thus the overall computation complexity of the ZFA is $O(P)$. This is much efficient than searching all possible collapsing window size, by which the computation complexity is $O(P!)$.

## 2.3 An alternative fast-Zoom method

The Zooming algorithm searches all binary partitions of a given region exhaustively. The computing speed of different statistical tests varies: some inherit probability distributions such as the W-test (Wang, *et al.*, 2016), while some incorporates permutations to obtain the p-values such as the SKAT (Wu, *et al.*, 2011). To assist the computation extensive methods to perform Zoom-Focus, we propose a fast-Zoom method: Instead of searching all possible partitions, the fast-Zoom performs a binary search, such that at each partition order $r$, the region is divided into two parts, only the part with smaller p-value is continued for the next level search, as illustrated in Diagram 2. The computation complexity of fast ZFA reduces to O(log2(P)).

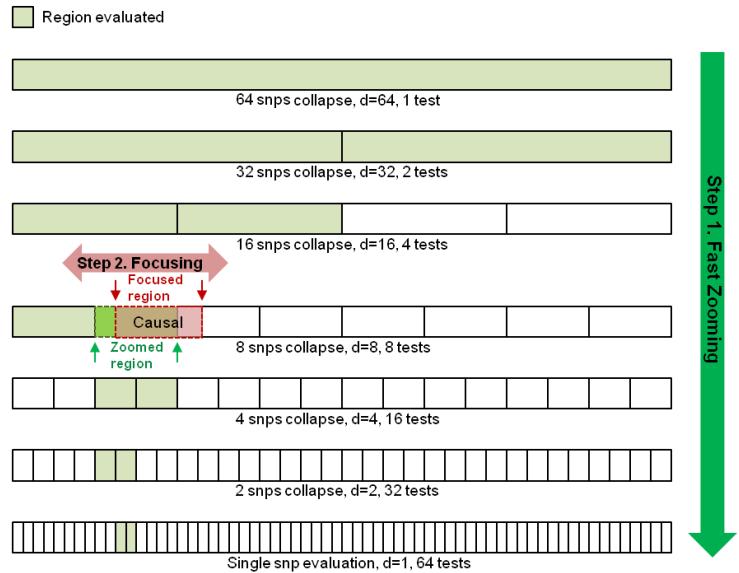

**Diagram 2.** Fast-Zoom-Focus Algorithm (fast-ZFA)

Legend: In fast-Zoom, a binary search replaces the exhaustive search - only green colored regions are evaluated. Fast-ZFA's computation complexity is O(log2($P$)), compared to ZFA's O($P$).

## 2.4 Simulation study design

Rare variant association test's power is largely influenced by the distribution of causal variants in the evaluation region (Sham and Purcell, 2014). To fully explore all methods' performance with ZFA, three scenarios are considered, with varying distributions and different effect directions (Diagram 3):

*Scenario I*: 8 causal SNPs cluster together in same effect direction
*Scenario II*: 8 causal SNPs cluster in two groups in same effect direction
*Scenario III*: 8 causal SNPs cluster together in opposite effect directions (5 SNPs show risk effect and 3 SNPs show protective effect)

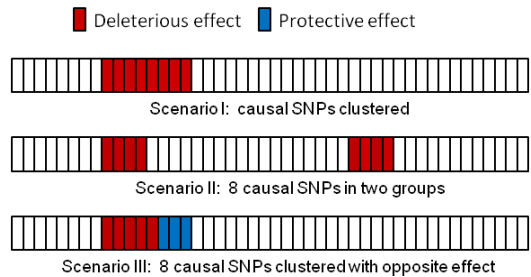

**Diagram 3.** Causal variants distribution scenarios in simulation study

To evaluate the effect of Focusing, distribution of causal variants are simulated to carry the setting of Scenario I but with uneven boundaries. Receiver Operating Curves (ROCs) are used to compare the Zooming and full ZFA performances.

Each simulated data contains 15,360 SNPs, 1,000 cases and 1,000 controls. The data is divided into 120 regions of P=128 SNPs. In each dataset, 10 regions are randomly selected to include causal variants according to one of the three scenarios, and the rest 110 regions are noise. The proportion of functional markers in a causal gene is 6.25% (8/128), and the proportion of causal genes in entire dataset is 8.33% (10/120). Phenotypes are generated using a logistic regression model of the causal SNPs (Wu, *et al.*, 2011):

$$LOGIT[\Pr(Y=1)] = \beta_0 + 0.5X_1 + 0.5X_2 + \sum_{i=1}^{8} \beta_i G_i,$$

where $X_1$ is a standard normal covariate, $X_2$ is a dichotomous covariate that takes the value 0 with probability 0.5 and the value 1 otherwise. $G_i$ are causal rare variants and $\beta_i$ are the effect size, $\beta_i=|\log_{10}MAF|\times 0.3$, such that rarer variants have larger effects. We controlled prevalence by $\beta_0$ and set it to 10% unless otherwise stated.

### 2.5 Statistical tests considered

The Zoom-Focus algorithm is applicable to all region-based rare variant association tests. Three representative methods are selected to combine with ZFA in the simulation studies, including the SKAT of variance component test; the WSS and the W-test of the burden test category. The SKAT is a quadratic score test that is composed of a weighted prediction error from a linear mixed model; it follows a mixture of Chi-squared distributions, and the p-value is obtained through Davies approximation (Wu, *et al.*, 2011). The SKAT is advantageous when a few large effect variants are located in a genomic region, and when the effect directions are not identical (Lee, *et al.*, 2014). The WSS uses Wilcoxon's rank sum statistic to measure a weighted count of the excess of mutations in the cases from the control, and use permutations to calculate p-values (Madsen and Browning, 2009). The WSS method is suitable for the scenario when the majority of causal variants have the same effect direction (Sha, *et al.*, 2012). The W-test is a fast and model-free genetic association test (Sun, *et al.*, 2016; Wang, *et al.*, 2016). It tests the distributional differences of SNPs in the case group from that in the control group, and follows a Chi-squared distribution with dataset adaptive degrees of freedom. For rare variant application, the W-test is calculated on the summed contingency table of SNPs from a given region (Sun, *et al.*, 2016). The WSS test and SKAT are computationally extensive and are applied with the fast-ZFA. The W-test is applied with both the fast and full ZFA.

### 2.6 Power, type I error calculation and ROC

One hundred replicated datasets are generated for each genetic scenario. In each replicate, 120 fixed regions are evaluated by alternative methods. A region is significant if it locates an optimal region with p-value smaller than $4.2 \times 10^{-4}$ (0.05/120), which is the Bonferroni corrected p-value at a family-wise error rate (FWER) of 0.05. An outcome is a true positive if the final optimal region overlaps the causal region. Power is the averaged true positive proportion from 100 simulated datasets, and the type I error rate is the average false positive rate of the 100 simulations with permuted phenotypes. Receiver Operating Characteristic (ROC) curves are used to assess the improvement of the Focusing step after the Zooming step.

**2.7 Real data application**

ZFA is applied on a real hypertensive disorder sequence data of the Genetics Analysis Workshop 19 (GAW19). The data consisted of 398 hypertensive patients and 1,453 healthy controls, and exome sequence of chromosome 3 is used. Variants with missing value percentage over 5%, MAF>1%, and inconsistent genotyping format are excluded (Laurie, *et al.*, 2010). After quality control, 41,788 rare variants are remained. Full ZFA is applied on Chromosome 3 using an initial window size $P = 256$. The chromosome is then divided into 163 non-overlapping regions, and the remaining 60 SNPs are grouped as the last window. To improve the computation efficiency, zoomed region with p-value smaller than $6.1 \times 10^{-4}$ (0.1/164) are passed to the Focusing step. A genetic region will be regarded as significant if its final p-value reach the Bonferroni corrected significance level of $3.0 \times 10^{-4}$ (0.05/164). Initial window sizes $P=128$ and $P=512$ are also applied, with significance threshold adjusted accordingly. Genes that include or overlap with the returned region are reported as susceptible.

## 3 RESULTS

**3.1 Simulation study**

Fast-Zoom and full path Zoom are evaluated under three causal marker distribution scenarios (Table 1); all rare variant tests received considerable power enhancement. In Scenario I, 8 causal SNPs cluster together with same effect direction. By fast-Zoom, the power of WSS and SKAT increased from below 2% to 58% and 19.3%, respectively; and W-test's power increased from 4.7% to 59.7%. For Scenario II, 4 causal SNPs form two distant clustering groups with the same effect direction. Before optimizing the collapsing region, SKAT gave the highest power of 25%, compared to WSS's 3.5% and W-test's 8.6%. After applying fast-Zoom, the power of SKAT increase to 70.0%, WSS increased to 74.8%, and the W-test increased to 71.3%. Both Scenario I and II favor unidirectional burden tests, thus WSS and W-test benefited most from optimizing the testing region. Especially in Scenario II, before optimization, SKAT had higher power because only a few causal SNPs are clustered, and it is disadvantageous for WSS and W-test when the testing region contains much noise; after Zooming, these two tests outperformed the SKAT. Scenario III contains 8 causal SNPs in different effect directions and is more suitable for variance component tests. Before optimization, SKAT showed the highest power of 11%, after optimizing the test region, SKAT's power increased to 50.5%, compared to WSS's 34.3% and W-test's 39.0%. Full-path Zoom with W-test gave the highest power under all scenarios; it gave the narrowest inter-quartile range (Q3 - Q1 =0) for all scenarios.

**Table 1.** Power and type I error rates of rare variant association tests before and after Zooming
Legend: Under all causal marker distribution scenarios, the statistical power of different rare variant association tests with Zooming shows a substantial enhancement comparing to without Zooming. The type I error rate is reasonably controlled.

| Scenarios of causal SNPs distribution | Statistical tests | Test Power before Zooming | Test Power after Zooming | Zoomed region size Median [Q1, Q3] |
|---|---|---|---|---|
| Scenario I[1] | WSS | 1.3% | 58.0% | 8 [8,16] |
| | SKAT | 1.8% | 19.3% | 8 [4,16] |
| | W-test | 4.7% | 59.7% | 8 [8,16] |
| | W-test* | 4.7% | 81.2% | 8 [8,8] |
| Scenario II | WSS | 3.5% | 74.8% | 8 [4,16] |
| | SKAT | 25.0% | 70.0% | 8 [4,16] |
| | W-test | 8.6% | 71.3% | 4 [4,8] |
| | W-test* | 8.6% | 94.2% | 4 [4,4] |
| Scenario III | WSS | 0.8% | 34.3% | 4 [4,8] |
| | SKAT | 11.0% | 50.5% | 8 [8,16] |
| | W-test | 2.0% | 39.0% | 4 [4,16] |
| | W-test* | 2.0% | 59.9% | 4 [4,4] |
| Type I error rate | WSS | 0.2% | 1.3% | 8 [2,16] |
| | SKAT | 0% | 0.1% | 4 [2,16] |
| | W-test | 0.9% | 4.8% | 8 [4,32] |
| | W-test* | 0.9% | 6.6% | 8 [4,32] |

[1] Scenario I: 8 unidirectional causal SNPs; Scenario II: Two clusters of 4 unidirectional causal SNPs; Scenario III: 8 bi-directional causal SNPs.
* Full-path Zoom, otherwise fast-Zoom

The fast-Zoom loses certain power compared to full path zooming, however, it is efficient and useful for tests that require permutation to generate p-values. The results showed that Zooming could generally locate the causal region and greatly improve rare variant testing power, either using the fast or full path. The type I errors are reasonably controlled (Table 1). For fast-Zoom, the SKAT's type I error rate was 0.1%, WSS's 1.3%, and the W-test's 4.8%. For full path Zoom, the type I error of W-test was 6.6%, reasonable in view of its power gain.

Figure 1 compared the ROC of full-path Zoom to the algorithm with Focusing step using W-test, simulated from scenario I. The ROCs showed that by applying the Zooming step alone, there was already considerable power, and the Focusing step further boosted the overall performance. The binary partition of Zooming resulted in discontinuous optimal region, and the true and false positive rates are returned in intervals, thus the ROC showed some form of a step function.

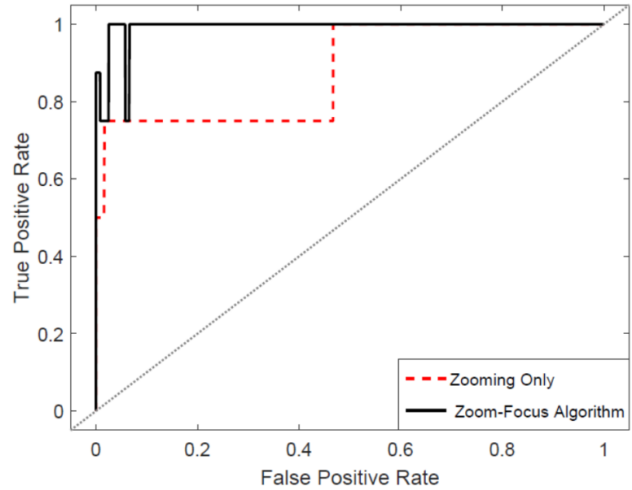

**Fig. 1.** Comparison of ROC curves of Zoom-only and the full Zoom-Focus Algorithm
Legend: The performance of rare variant test can be further improved by the Focusing step.

### 3.2 Computation time

On a PC cluster with 8 compute nodes each with 16GB RAM (2GB per core) and 2.66GHz E5430 CPU, the unit time elapsed for fast-Zoom in a region of 128 SNPs is 97.63 seconds(s) by WSS, 295.57s by SKAT, and 0.18s by W-test (Table 2). ZFA with W-test is several hundred times faster than with WSS and SKAT. Full path Zoom by the W-test took 2.28s. In real data, full ZFA with W-test took approximately 15 minutes to complete the whole exome evaluation of chromosome 3.

**Table 2.** Computing time* comparison of different methods

|  | WSS | SKAT | W-test |
|---|---|---|---|
| Fast-Zoom | 97.63 seconds | 295.57 seconds | 0.18 seconds |
| Zoom | \ | \ | 2.28 seconds |

* Note: The computing time is calculated by cluster of 8 cores

### 3.3 Real data application

On whole exome sequencing of chromosome 3 of hypertensive disorder, ZFA with W-test identified two significant regions, one contains 29 SNPs and the other has 27 SNPs (Table 3). Different initial window sizes are applied and gave the same optimal region (Supplementary Materials S1). This showed that the method is quite robust. The two regions are parts of the genes *ITPR1* and *MSL2/ PCCB* (http://www.ncbi.nlm.nih.gov). Both genes were previously reported to associate with cardiovascular and metabolic traits (Consortium, 2013; Dehghan, *et al*., 2009; Kathiresan, *et al*., 2007; Sabater-Lleal, *et al*., 2013). The gene *ITPR1* spans 251 SNPs, and *MSL2* covers 85 SNPs (Table 3). We also calculated the p-values of these two genes using gene-based WSS, SKAT and W-test without region optimization, and none of them are significant) (Figure 2). The results showed that ZFA could greatly enhance the effective selection of rare variants with disease association on exome sequencing data. The phenotypes are also permuted for 100 times to evaluate possible false positives, and the average false positive rate is 1.7%, which indicated that there is no inflation of spurious association.

**Table 3.** Significant genes detected by ZFA and W-test on chromosome 3 of real hypertensive disorder data

| Chr | Gene | Total number of SNPs in gene | ZFA optimized testing region size | P-value of identified region |
| --- | --- | --- | --- | --- |
| 3 | *ITPR1* | 251 | 29 | 1.99E-04 |
| 3 | *MSL2/PCCB* | 85 | 27 | 3.42E-07 |

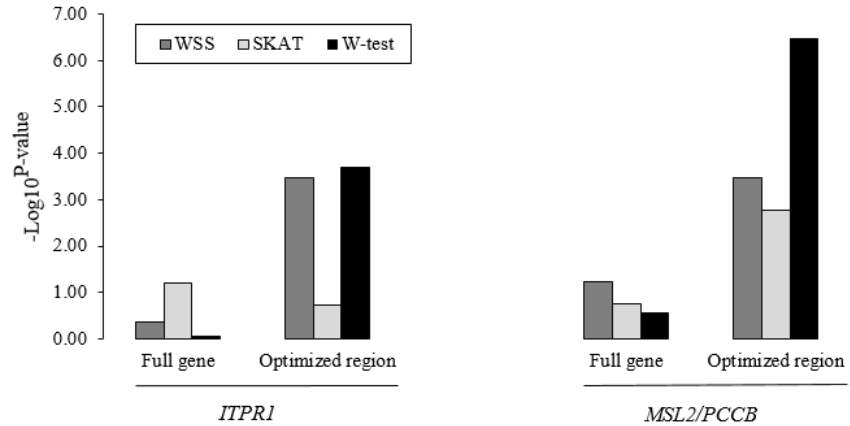

**Fig. 2.** P-value of genes with and without ZFA optimization

Legend: ZFA with W-test identified two significant genes on exome of chromosome 3 of hypertensive subjects. Both genes are insignificant by gene-based tests without testing region optimization.

## 4 DISCUSSION

The search for rare variant's contribution of complex disease was triggered by the hypothesis that a few SNPs may account for large effect sizes (Manolio, *et al*., 2009). However, increasing evidences have shown that rare or low frequency genetic variants have modest-to-weak effect sizes (Auer and Lettre, 2015). These findings suggest that for rare variant association study, it is especially important to consider the effect of multiple loci jointly. Furthermore, functional annotation of rare variants is still evolving; there are multiple platforms to determine the actual regions of genes, some with overlapping boundaries (2012; Kircher, *et al*., 2014; Maurano, *et al*., 2012). There is also no consensus of whether the promoters, un-translated regions or introns should be included in gene-base tests (Auer and Lettre, 2015). Even for the coding variants, it is likely that only a part of them is functional and others represent random genetic variation (Ionita-Laza, *et al*., 2014). The problem is more serious with the increasing sequencing depth – a common gene easily spans two to three thousand SNPs. The power of direct application of aggregation tests would be affected from the large proportion of possible noises, thus it is crucial to determine the region of testing as a starting point.

The motivation of ZFA is to locate an optimal region with maximum association information and to exclude noise, based on the information of the data. In fact, the Zooming step is performing a feature selection within a given genomic region, and the features are SNP-clusters of varying sizes. Based on the best partition, boundaries are further refined by adding or subtracting adjacent SNPs surrounding them. In this way, noise can be discarded and statistical power can be improved. Simulation studies demonstrated that the ZFA can boost power of various rare variant tests by over ten-fold.

In real data analysis of one chromosome by ZFA, only two regions are significant and they belong to the protein coding genes *ITPR1* and *MSL2/PCCB*. Both genes were previously found to have association with cardiovascular and metabolic disorders in independent studies: the *ITPR1* was reported in the Framingham Heart Study to be associated with cholesterol levels (Kathiresan, *et al*., 2007). The *MSL2*/*PCCB* was found to have a strong association with fibrinogen levels, coronary artery disease, blood pressure, and body mass index (Consortium, 2013; Dehghan, *et al*., 2009; Sabater-Lleal, *et al*., 2013). In this dataset, neither of the two genes was significant by gene-based tests, but they are identified through optimizing the testing region. Specifically, the *ITPR1* encodes an intracellular receptor for inositol 1,4,5-trisphosphate. It plays a role in mediating calcium release from the endoplasmic reticulum. The *MSL2* is a subunit of a protein complex that functions in chromatin modification, and the protein encoded by *PCCB* is a subunit of the propionyl-CoA carboxylase (PCC) enzyme. Application of ZFA with rare variant aggregation tests increased the chance of identifying disease susceptible loci. More experiments are needed to confirm the discovered regions' underlying biological functions related to disease.

To conclude, we propose a Zoom-Focus Algorithm (ZFA) as an efficient, robust and powerful approach to enhance statistical power of rare variant association tests by locating the optimal testing region. The algorithm is flexible to be applied with various rare variant methods and scalable for whole genome analysis. It is a practical and powerful approach in exome sequencing data analysis towards elucidating the role of rare variant in complex disorder.

**ACKNOWLEDGEMENTS**

This study makes use of data provided by the Genetic Analysis Workshop 19 (GAW19), funded by the National Institutes of Health through grant R01 GM031575.

*Funding*: This work is partly supported by the Chinese University of Hong Kong Direct Grant [4054169 to MHW]; the Research Grant Council – General Research Fund [476013 to MHW]; and the National Science Foundation of China [81473035, 31401124 to MHW].

Conflict of Interest: none declared.

# Supplementary Materials

**S1 Regions detected with different initial window sizes in real data application**

When the initial window size P = 128, chromosome 3 is divided into 326 non-overlapping region, and the remaining 60 SNPs are grouped as the last window. Zoomed region with p-value smaller than $3.1 \times 10^{-4}$ (0.1/327) are passed to the Focusing step. A genetic region will be regarded as significant if its final p-value reach the Bonferroni corrected significance level of $1.5 \times 10^{-4}$ (0.05/327).

**Table S1 (A)**. Significant genes detected on chromosome 3 with initial window size P = 128

| Chr | Gene | Total number of SNPs in gene | ZFA optimized testing region size | P-value of identified region |
|---|---|---|---|---|
| 3 | *ITPR1* | 251 | 29 | 1.11E-04 |
| 3 | *MSL2/PCCB* | 85 | 27 | 1.71E-07 |

When the initial window size P = 512, chromosome 3 is divided into 81 non-overlapping region, and the remaining 316 SNPs are grouped as the last window. Zoomed region with p-value smaller than $1.2 \times 10^{-3}$ (0.1/82) are passed to the Focusing step. A genetic region will be regarded as significant if its final p-value reach the Bonferroni corrected significance level of $6.1 \times 10^{-4}$ (0.05/82).

**Table S1 (B)**. Significant genes detected on chromosome 3 with initial window size P = 512

| Chr | Gene | Total number of SNPs in gene | ZFA optimized testing region size | P-value of identified region |
|---|---|---|---|---|
| 3 | *ITPR1* | 251 | 29 | 3.98E-04 |
| 3 | *MSL2/PCCB* | 85 | 27 | 6.49E-07 |